\documentclass[useAMS,usenatbib,referee]{mn2e}
\usepackage{graphicx}
\usepackage{amssymb}
\usepackage[T1]{fontenc}
\usepackage{subfig}
\usepackage{color}

\title[The FEROS--Lick/SDSS  database of spectral indices]{The FEROS--Lick/SDSS observational database of 
spectral indices of FGK stars for stellar population studies}
\author[ M. Franchini, C. Morossi, P. Di Marcantonio, M. L. Malagnini, and M. Chavez ]{M. Franchini$^{1}$\thanks{E-mail:
franchini@oats.inaf.it (MF)}, C. Morossi$^{1}$, P. Di Marcantonio$^{1}$, 
M.L. Malagnini$^{1}$, and M. Chavez$^{2}$\thanks{E-mail: mchavez@inaoep.mx (MC)}\\
$^{1}$INAF - Osservatorio Astronomico di Trieste, Via G. B. Tiepolo 11, Trieste, I-34143, Italy\\
$^{2}$Instituto Nacional de Astrof\'isica, \'Optica y Electr\'onica, Luis Enrique Erro 1, 72840 Tonantzintla, Puebla, Mexico}

\begin{document}

\date{Accepted 2014 April 28.  Received 2014 April 28; in original form 2014 April 14 }

\pagerange{\pageref{firstpage}--\pageref{lastpage}} \pubyear{2014}

\maketitle

\label{firstpage}

\begin{abstract}

We present FEROS--Lick/SDSS, an empirical database of  Lick/SDSS spectral indices of FGK stars to be used  in  
population synthesis projects for discriminating different stellar 
populations within the integrated light of galaxies and globular clusters. From about 2500 FEROS stellar spectra obtained from the ESO Science Archive Facility
we computed line--strength indices for 1085 non--supergiant stars with atmospheric parameter estimates from the AMBRE project. 

Two samples of 312 {\it dwarfs} and of 83 {\it subgiants} with solar chemical composition and no significant $\alpha$--element abundance enhancement
are used to compare their observational indices with the predictions of the Lick/SDSS library of synthetic indices. 
In general, the synthetic library reproduces very well the behaviour of observational indices as a function of temperature, but in the case of low temperature 
($T_{\rm eff}$ $\lesssim $5000\,K) dwarfs; low temperature subgiants are not numerous enough to derive any conclusion. 
Several possible causes of the disagreement are discussed and promising theoretical improvements are presented.

\end{abstract}


\section{Introduction}
Spectra of galaxies and globular clusters carry a wealth of information about gas and different stellar populations properties. 
The analysis of stellar populations is of primary importance for the understanding of the physical  processes involved 
in the formation and evolution of galaxies because it provides a unique tool to evaluate metal enrichment and star formation epoch(s). 

There are several approaches to get information about abundance patterns in stellar populations 
that use colours (e.g. \citealt{JA06}; \citealt{CA09}), broad and narrow spectral features or indices (e.g. \citealt{RO94};
\citealt{WR94}; \citealt{CE09}; \citealt{WR14b}; \citealt{VA10}; \citealt{SA13}), or  full spectral fitting
(e.g. \citealt{WA09}). Colours are still useful in studying faint objects for which spectral features may not be obtainable, 
but  are strongly affected by dust extinction.

Full spectral fitting and spectral indices, on the other hand,  are preferred when studying brighter objects because
they allow studies of individual chemical species. These spectroscopic studies involve a process that compares
model predictions with observed spectra for the brightest objects or with line intensities of the most prominent atomic and molecular absorptions
for fainter objects. Additionally, the latter, unlike broad band colours and full spectra, are almost unaffected by interstellar reddening (e.g.  \citealt{MA05})
and by spectral energy calibration uncertainties.

For a long time, spectroscopic analyses of stellar populations have relied on the Lick/IDS system of indices (\citealt{GO93}; 
\citealt{WR94}; \citealt{WR97}). It is important to 
remark, however, that the Lick/IDS system was defined at a resolution (R$\sim$630)
which is much lower than the ones  of recent and forthcoming surveys (like, for example, the
Sloan Digital Sky Survey, SDSS \citealt{YOR00}, and the Large Sky Area Multi--Object Fiber 
Spectroscopic Telescope survey, LAMOST\footnote{http://www.lamost.org/LAMOST}).
Furthermore, as discussed in \citet{WR94}, several uncertainties in both IDS response
function and wavelength calibration make it difficult to transform new observational indices
into the original Lick/IDS system thus introducing possible systematic errors even if several
improvements with respect to the past are now possible by using \cite{WR14a}. 

At times when there is an easy access to huge high quality spectral surveys, it is of fundamental importance to test, evaluate and 
eventually improve the current theoretical machinery, to keep pace  with the fast observational development.

\cite{FR10} presented a new synthetic library, the Lick/SDSS library,
of indices in a Lick--like system fine tuned to analyze data at medium resolution like those of SDSS and LAMOST.
The use of R=1800 (Sloan/SDSS resolution) allows us to avoid a potential loss 
of information that would occur in degrading SDSS spectra. Furthermore, 
the Lick/SDSS system was built from flux calibrated
spectral energy distributions and therefore is not characterized by any instrumental signature.

The Lick/SDSS library is primarily aimed at applications in stellar population synthesis, specifically 
to the study of old and intermediate--age stellar populations, but it also represents
a useful tool for determining F,G, and K stellar  atmospheric parameters and abundance ratios, in particular 
[Ca/Fe] and [Mg/Fe],  \citep{FR11}. In fact, the study of abundance patterns like,
for example, $\alpha$--element enhancement, gives insight into the role of SNe\,I and SNe\,II
in the chemical enrichment of galaxies. Moreover, the library can be used to complement empirical libraries
in segments of the stellar parameter space, metallicity in particular, 
that are not homogeneously covered by observations. In fact, in general, empirical libraries are
built with objects that carry on the imprints of the local properties 
of the solar neighbourhood, hence, might not be applicable to study 
the integrated spectra of stellar systems with different star formation histories, such as elliptical galaxies.
It is fair to mention that synthetic libraries do not suffer from
the above mentioned limitation but, on the other hand, the synthetic approach is prone to 
uncertainties that are mostly related to the approximations
associated with theoretical model atmospheres and with the completeness of the lines lists
used in computing synthetic spectra.
It is therefore mandatory, before blindly using libraries of synthetic 
indices, to perform exhaustive checks on how well the theoretical 
predictions match the available observations of real stars, and to
quantitatively establish their applicability and limitations.

Along this line, in this paper we produce an empirical library of Lick/SDSS indices for 1085  non--supergiant stars
obtained by using  observations taken with the Fibered Extended Range Optical Spectrograph 
(FEROS) \citep{KA99}.  In Section\,\ref{obs}  we 
describe the sample of the analyzed stars and the computation of their observational indices,
and present the obtained FEROS--Lick/SDSS database. 
Eventually, in Section\,\ref{result} we compare the observational indices of two sub--samples of 312 dwarfs and of 83 
subgiants of solar chemical composition (-0.2$<$[M/H]$<$+0.2)  with the predictions of the Lick/SDSS library
and discuss the results.

\begin{figure*}
       \subfloat[Distributions of $T_{\rm eff}$, log\,$g$, metallicity and $\alpha$--enhancement  \label{param1}]{%
      \includegraphics[width=0.6\textwidth]{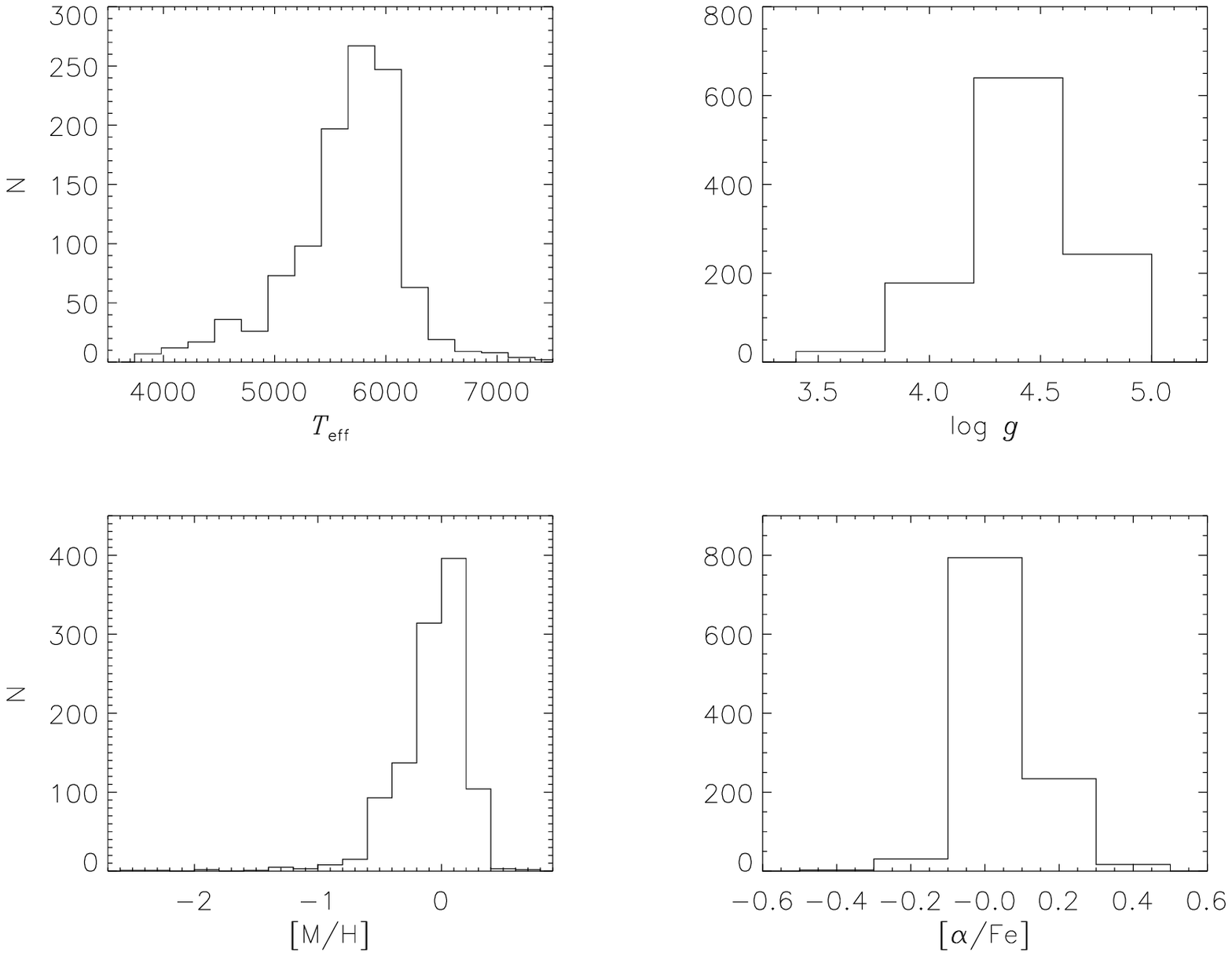}
    }
    \hfill
    \subfloat[log\,$g$--log\,$T_{\rm eff}$ diagrams\label{param2}]{%
      \includegraphics[width=0.3\textwidth]{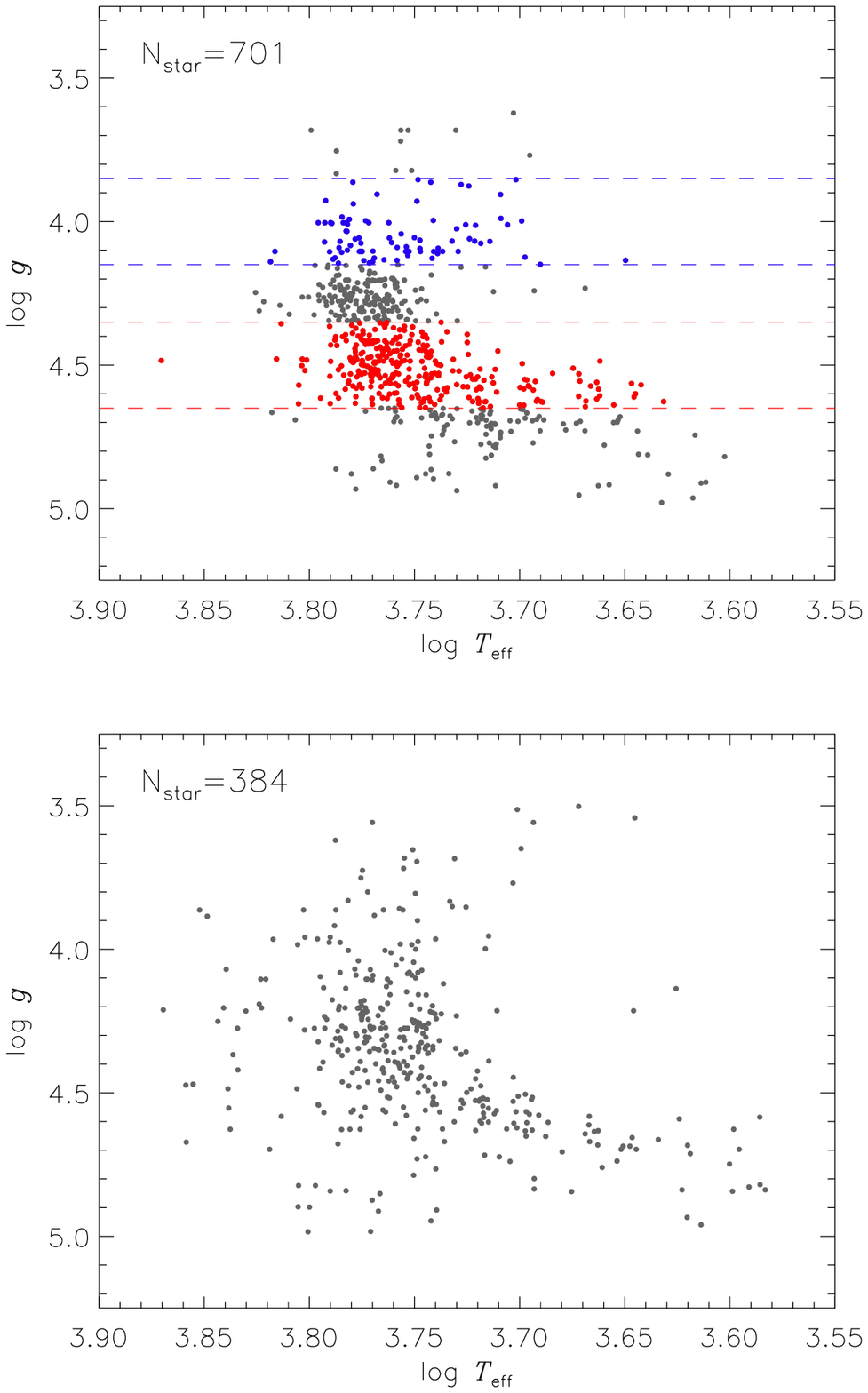}
    }
    \caption{Distributions of $T_{\rm eff}$, log\,$g$, [M/H], and [$\alpha$/Fe] and log\,$g$--log\,$T_{\rm eff}$ diagrams (upper panel: solar chemical composition stars, see text
     for description of red and blue points;
     lower panel: remaining stars) for the complete sample of 1085 stars.}
    \label{param}
\end{figure*}

\section{Program stars and observational FEROS--Lick/SDSS indices}
\label{obs}
In order to build an empirical database of observational indices, we need  a set  of stars with reliable estimates of 
atmospheric parameters (i.e. effective temperature ($T_{\rm eff}$), surface gravity (log\,$g$), and global metallicity ([M/H])) 
from whose spectra observational indices can be computed. After such a collection has been assembled, the comparison between 
observational indices and predictions of theoretical libraries built from model atmospheres and synthetic spectra 
is  straightforward  by imposing the match of the stellar and model atmospheric parameter values. 

The stars observed by FEROS  and studied by the AMBRE project 
\citep{WOR12} constitute an ideal working dataset for our purposes since they include a large number of non--supergiant FGK stars with
individual estimates of  $T_{\rm eff}$, log\,$g$, [M/H], and $\alpha$ to iron ratio ([$\alpha$/Fe]).

We searched the ESO Science Archive Facility and retrieved, through the FEROS/HARPS pipeline processed data Query 
Form\footnote{http://archive.eso.org/wdb/wdb/eso/repro/form}, all the public available spectra of FGK stars 
with  AMBRE  atmospheric parameter values in the following ranges:
$3800 < T_{\rm eff} < 7500$\,K, ${\rm log} g > 3.5$, and global metallicity  [M/H]$ > -3.0$.

A list of 1085 stars, corresponding to 2511 available spectra, was obtained. Since AMBRE
provides individual estimates of stellar parameters derived from each spectrum, we computed for  202 stars with more than one observed spectrum
average atmospheric parameter values. In any case the dispersion of values for the same object resulted to be less than the external errors
associated with AMBRE results. 

Figure\,\ref{param1} shows the distributions of  $T_{\rm eff}$, log\,$g$, [M/H], and  [$\alpha$/Fe] values for the 1085 stars{\footnote{Only 
1079 stars have [$\alpha$/Fe] estimates} built by using half--bin sizes corresponding to the external errors given in AMBRE 
(i.e. $\Delta T_{\rm eff}$=120\,K,  $\Delta$log\,$g$=0.2\,dex, $\Delta$[M/H]=0.1\,dex, 
and $\Delta$[$\alpha$/Fe]=0.1\,dex). As can be seen, most of the stars have dwarf or subgiant surface gravities,
solar metallicity, and  no significant $\alpha$--element abundance enhancement. In Fig.\,\ref{param2} we show,  in the upper panel, the 
log\,$g$--log\,$T_{\rm eff}$ diagrams for the 701 solar metallicity stars (-0.2$<$[M/H]$<$+0.2 and -0.2$<$[$\alpha$/Fe]$<$+0.2) while the other  384 stars   
are shown in the lower panel.

In order to compute observational Lick/SDSS indices, all 2511 spectra were corrected for radial velocity adopting the values given in AMBRE, 
and, subsequently, degraded to the resolution of the Lick/SDSS library ($R$=1800).
 A set of FEROS indices were computed for each spectrum and averaged in the case of multiple observations of the same
star. An additional step is required to transform FEROS indices into the Lick/SDSS system since FEROS spectra are not flux calibrated  and thus can be impacted to a certain 
extent by the instrumental response. In order to perform such a transformation,
we looked for FEROS stars in common with those contained in at least one of the databases used in \citet{FR10} to calibrate the Lick/SDSS library, namely,
ELODIE \citep{MO04}, INDO--U.S.\citep{VA04}, and MILES \citep{CE07}.

Figure\,\ref{calib} shows some examples of the comparison between FEROS and reference Lick/SDSS values from the 
above mentioned databases for the 58 FEROS stars in common. 
In general, the regression lines present slopes very close to one and small systematic offsets. Exceptions are the CaHK index,
whose index definition wavelength interval falls in the blue region of FEROS spectra where the SNR is, in general, low, and  the Mg$_1$ and  Mg$_2$ indices 
which  span such a large wavelength interval to make  the FEROS
instrumental response variation quite important. The scatter around the regression lines 
is always consistent with the typical errors given in \cite{FR10} 
thus showing the reliability of the transformation of FEROS indices into the Lick/SDSS system.

 The set of transformation coefficients of the linear regressions such as those shown in Fig.\,\ref{calib} were used to transform the 
FEROS index values into the Lick/SDSS system thus  obtaining the FEROS--Lick/SDSS library, i.e.  the empirical database of FEROS--Lick/SDSS index 
values of 1085 FGK stars, which is presented 
in Table\,\ref{tab1}.  For each star we provide the number of FEROS spectra, $N_{\rm sp}$, the mean atmospheric parameter values together with
their standard deviations, and the mean index values with the corresponding standard deviations.

\begin{figure*}
\includegraphics[width=160mm]{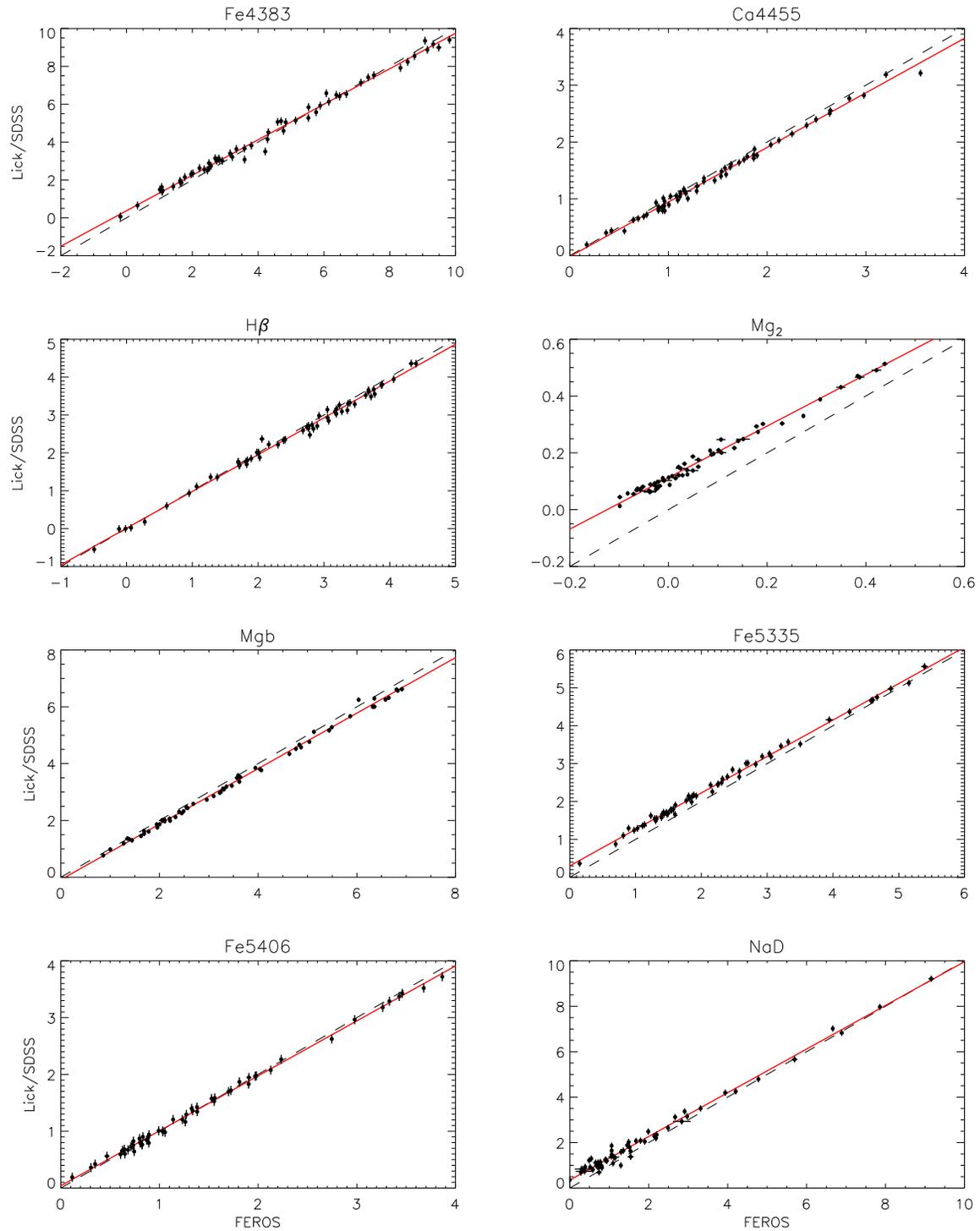}
\caption{Examples of transformation of FEROS indices into the  Lick/SDSS system.  In each panel we show Lick/SDSS index values by
\citet{FR10} versus FEROS index values, together with the 45$^{\rm o}$ line (black dashed) and the regression line (red solid). }
\label{calib}
\end{figure*}

\begin{table*}
\tiny
 \centering
  \caption{Table of observational Lick/SDSS indices of 1085 FGK FEROS stars.The atmospheric parameters  $T_{\rm eff}$, log\,$g$, [M/H], and  [$\alpha$/Fe] and index values
are averaged values in the case of more than one spectrum  (see text).  All the 19 Lick/SDSS indices are available in the on-line version.}
\label{tab1}
  \begin{tabular}{@{}lccrccrccccccc@{}}
  \hline
   \multicolumn{1}{c}{Name} & $N_{\rm sp}$ &  $T_{\rm eff}$  &  \multicolumn{1}{c}{$\sigma_{T_{\rm eff}}$}  &  log\,$g$ & $\sigma_{{\rm log} g}$ &  \multicolumn{1}{c}{[M/H]} &     
   \multicolumn{1}{c}{$\sigma_{{\rm [M/H]}}$}    & [$\alpha$/Fe]  & $\sigma_{[\alpha/{\rm Fe}]}$ & CaHK & $\sigma_{{\rm CaHK}}$ & CN$_1$ &  $\sigma_{{\rm CN}_1}$  \\
       &   & K &  \multicolumn{1}{c}{K}  &  & dex &   & dex & & dex & \AA{} & \AA{}  & mag & mag \\
  \hline
       HD 224725  &    1  &      6319  &        38  &       4.98  &       0.16  &      -0.38  &       0.08  &       0.07  &       0.05  &      8.901     &      ---   &      -0.044  &   --- \\
        HD 224810  &    1  &      5738  &        12  &       4.30  &       0.10  &       0.06  &       0.04  &       0.02  &       0.02  &       2.919   &      ---  &   -0.027  &   --- \\
       HD 224828  &    1  &      5598  &        12  &       4.45  &       0.04  &      -0.47  &       0.02  &       0.16  &       0.01  &     7.591   &      ---  &       -0.084  &   --- \\
        HIP 112  &    3  &      3970  &         7  &       4.84  &       0.01  &       0.27  &       0.03  &      -0.30  &       0.01  &     13.000    &      3.693  &      ---  &   --- \\
       ...  &   ...  &   ...  &       ...     &   ...     &   ...      &    ...    &     ...    &   ...     &    ...   &     ...   &    ...     &   ...      &    ...\\
 
\hline
\end{tabular}
\end{table*}

\begin{figure*}
\includegraphics[]{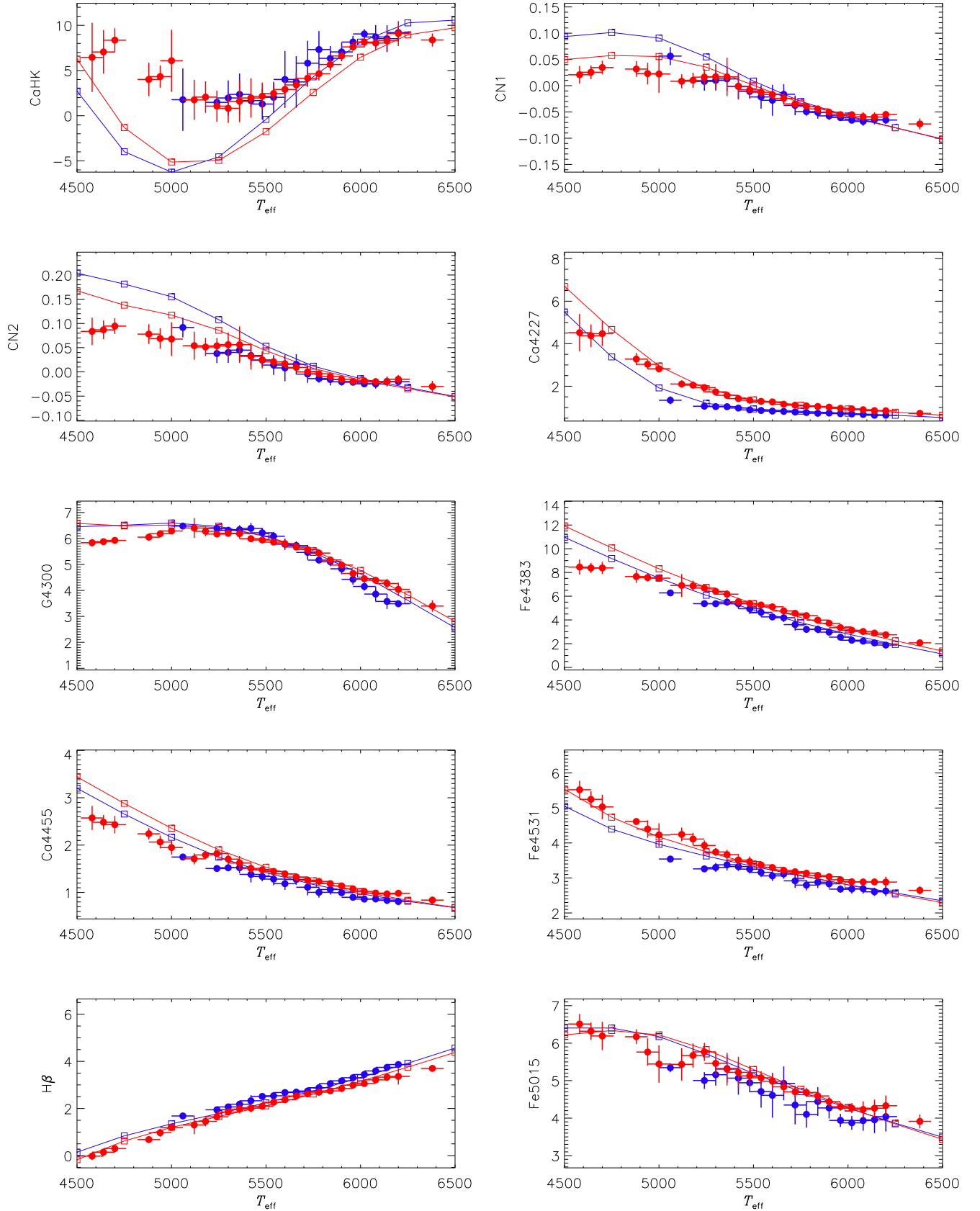}
\caption{FEROS--Lick/SDSS indices versus  $T_{\rm eff}$ for {\it dwarfs} (red points) and  {\it subgiants} (blue points). Empirical index values 
re-binned at half--overlapped intervals of amplitude equal to AMBRE external  $T_{\rm eff}$ errors (120\,K) are compared with the predictions of the Lick/SDSS library 
at [Fe/H]=+0.0, [$\alpha$/Fe]=+0.0, and log\,$g$=4.0 (blue lines) or log\,$g$=4.5 (red lines). }
\label{teff}
\end{figure*}

\begin{figure*}
\includegraphics[]{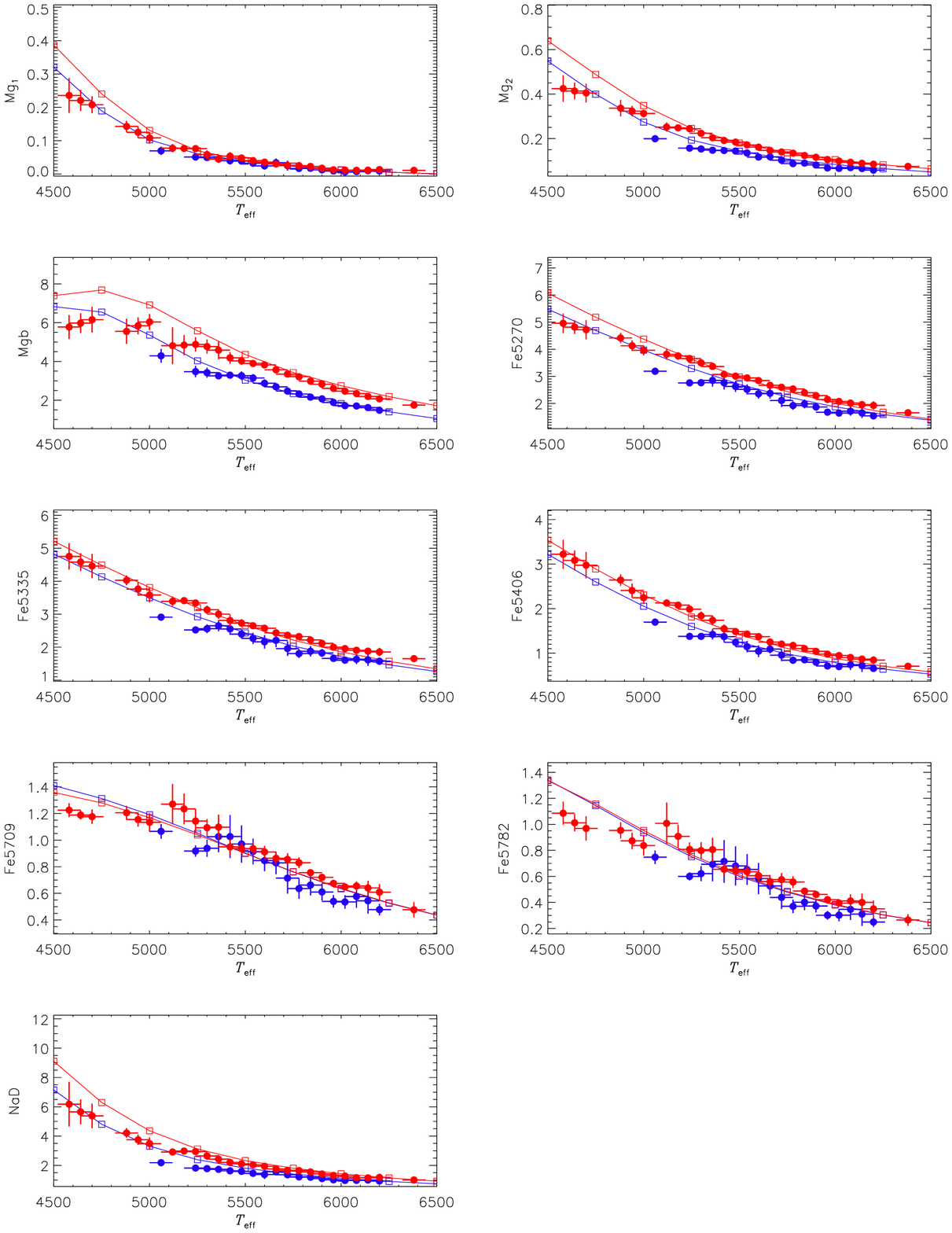}
\contcaption{ }
\end{figure*}

\section{Comparison with the Lick/SDSS library}
\label{result}
In order to check the  ability of synthetic Lick/SDSS indices to reproduce the observations of non--supergiant
FGK stars, we can use the derived empirical FEROS--Lick/SDSS database presented in the previous section.
Statistically sound results can be obtained only for solar chemical composition dwarfs and subgiants due to the paucity and inhomogeneous distribution of the
other stars in the parameter space. 
Therefore, we extracted  out of the 1085 FEROS stars the following two groups of objects:
\begin{enumerate}
\renewcommand{\theenumi}{(\arabic{enumi})}
 \item {\it subgiants} (log\,$g\simeq$4.0):  this first group consists of 83 stars (blue points in the top panel of Fig.\,\ref{param2})
 with the atmospheric parameter values given in Table\,\ref{tab1} in the following ranges:
$ 4450 < T_{\rm eff} < 6600$\,K, $3.85 < {\rm log} g < 4.15$, $-0.2 < $[M/H]$ < +0.2$, and  $-0.2 < $[$\alpha$/Fe]$ < +0.2$;
\item {\it dwarfs} (log\,$g\simeq$4.5): this second group consists of 312 stars (red points in the top panel of Fig.\,\ref{param2}) 
with the atmospheric parameter values given in Table\,\ref{tab1} in the following ranges:
  $4250 < T_{\rm eff} < 7500$\,K,  $4.355 < {\rm log} g < 4.65$,  $-0.2 < $[M/H]$ < +0.2$, and  $-0.2 < $[$\alpha$/Fe]$ < +0.2$.
\end{enumerate}

The intervals in log\,$g$, [M/H], and [$\alpha$/Fe] were chosen taking into account the corresponding external errors given in AMBRE in order to obtain two practically
non--overlapping groups to be compared with the Lick/SDSS library predictions for [M/H]=0.0, [$\alpha$/Fe]=0.0, and log\,$g$=4.0 and log\,$g$=4.5, respectively.

Figure\,\ref{teff} shows the re--binned observational FEROS--Lick/SDSS index values versus  $T_{\rm eff}$ for the {\it subgiants} and {\it dwarfs}
superimposed onto the predictions  of the Lick/SDSS library. There is a general good agreement of {\it dwarfs} (red points) and {\it subgiants}
(blue points) with the  log\,$g$=4.5 and log\,$g$=4.0 lines (red and blue, respectively) for  $T_{\rm eff}\gtrsim 5000$\,K. 

In the following we describe in more detail the comparison for each individual index:
\begin{description}
 \item CaHK: The observational values for this index are always higher than the predicted ones for the  {\it dwarfs} with the only exception of the
higher temperatures ($T_{\rm eff}> $ 6000\,K); there is a good agreement for the {\it subgiants} for $T_{\rm eff}> $ 5500\,K. We recall that this index
fall in the extreme blue region of FEROS spectra which is, in general, characterized by a low SNR, and where the transformation of empirical FEROS
indices into the Lick/SDSS system is more critical. Nevertheless, we cannot be sure that the disagreement between observational and
synthetic values can be totally due to these observational problems;
 \item CN$_1$, CN$_2$: The predictions of the Lick/SDSS library match the observational indices for  {\it dwarfs} for  $T_{\rm eff}> $ 5300\,K for both indices,
while for the {\it subgiants} a good agreement is achieved for $T_{\rm eff}> $ 5600\,K only. These indices are very sensitive to the abundances of C, N, and O
and no information about these elements in AMBRE;
 \item Ca4227: Observational and synthetic index values agree for  {\it dwarfs} for  $T_{\rm eff}> $ 4750\,K; {\it subgiants} data are available only for 
  $T_{\rm eff}> $ 5000\,K and show hints of an overestimation of the index by the Lick/SDSS library below 5250\,K; 
 \item G4300: There is a very good agreement for {\it dwarfs} for all temperatures above 5100\,K; discrepancies can be seen for  {\it dwarfs} for $T_{\rm eff}< $ 5000\,K
and for {\it subgiants} for  $T_{\rm eff}> $ 6000\,K. This index, as CN$_1$, CN$_2$ may be affected by uncertainties in Carbon abundance.
 \item Fe4383: Observational and synthetic index values agree very well for  {\it dwarfs} for  $T_{\rm eff}> $ 5100\,K and for  {\it subgiants} for  $T_{\rm eff}> $ 5400\,K. 
The synthetic index values  overestimate the observational ones for the coolest  {\it dwarfs};
 \item Ca4455: This index behaves as Fe4383;
 \item Fe4531: The Lick/SDSS library predictions reproduce very well the values of  observational indices for {\it dwarfs} for all temperatures and for  {\it subgiants}
for $T_{\rm eff}> $ 5300\,K.  Synthetic index values are larger than the observational ones for the coolest   {\it subgiants};
 \item H$\beta$: There is a very good agreement both for {\it dwarfs} and   {\it subgiants} for all temperatures. It is worth stressing the 
goodness of this result since, for this index,  significant discrepancies between theoretical predictions and observations in the Lick/IDS system 
are shown in Fig.\,1 by \cite{WR14b};
 \item Fe5015: This index shows a quite large observational scatter. The synthetic index values are larger than the observational ones for  {\it subgiants}
and for {\it dwarfs} for 4900 $< T_{\rm eff}< $ 5200\,K;  
  \item Mg$_1$, Mg$_2$, and  Mgb:  Observational values for  {\it dwarfs} and {\it subgiants} are very well in agreement with the Lick/SDSS library predictions
for $T_{\rm eff}> $ 5250\,K, for Mg$_1$ and Mg$_2$,  and  for $T_{\rm eff}> $ 5500\,K for  Mgb. At lower temperatures the observational indices fall below the
predicted ones. It is worth noticing that no agreement was achieved between empirical and synthetic Mgb values in the Lick/IDS system (see Fig.\,1, by \citealt{WR14b}); 
  \item Fe5270:  There is a very good agreement for {\it dwarfs} for all temperatures above 5100\,K and for  {\it subgiants} for  $T_{\rm eff}> $ 5400\,K. In this
case, the agreement between theory and observations is much better than in the Lick/IDS system (see again  Fig.\,1, by \citealt{WR14b}). Moreover,
the improvement in sensitivity
to surface gravity of the Lick/SDSS Fe5270 index with respect to the Lick/IDS one is also evident since  {\it subgiant} and {\it dwarf} observational sequences 
are well separated in the diagram;
 \item Fe5335 and Fe5406: There is a very good agreement for {\it dwarfs} for all temperatures while for {\it subgiants}  this index behaves as Fe5270; 
 \item Fe5709 and Fe5782: These two indices are almost insensitive to surface gravity; the observational index values show a quite large scatter and fall
below the Lick/SDSS library predictions for  $T_{\rm eff}< $ 4750\,K for {\it dwarfs};
 \item NaD: The synthetic index values overestimate the observational ones in particular for the coolest  {\it dwarfs} for $T_{\rm eff}< $ 5250\,K.
\end{description}

In conclusion, there is a  generally good agreement between theoretical predictions and observational index values for temperatures above $\sim$\,5250\,K 
showing that the Lick/SDSS library can be safely used to complement empirical databases for relatively ``hot'' FGK stars.
 At lower temperatures there is a reasonable agreement for Ca4227, Fe4531, H$\beta$,  Fe5335, and Fe5406  for {\it dwarfs} while the paucity of ``cool'' {\it subgiants}
prevent us to draw a sound conclusion for this kind of stars. 
In all the other cases the Lick/SDSS library predicts too strong index values for temperatures below 5250\,K.

An apparent disagreement between the Lick/SDSS library synthetic indices and observed ones was already
found by \cite{FR10} by using SDSS--DR7 stellar spectra. This disagreement was explained by systematic offsets in the stellar $T_{\rm eff}$  estimates
derived by using SEGUE Stellar Parameter Pipeline \citep{LE08} with respect to the temperature scale of the Lick/SDSS library\footnote{The persistence of 
these systematic offsets after the changes and improvements made on the SSPP and  described  in \citet{SM11} should be investigated.}.  
A similar explanation for the discrepancies found in this work seems untenable, since Fig.\,\ref{teff} shows a good match for the
the most $T_{\rm eff}$ sensitive index, i.e. H$\beta$. A further
indication that the problem is not in the AMBRE  temperature determinations is given by Fig.\,\ref{ind} where we compare
predicted  and observational indices in index--index diagrams which are independent of temperature determinations. As can be seen,
the indices of most of the stars with AMBRE  temperature estimates below 5250\,K (right panels in Fig.\,\ref{ind}) 
fall in regions which are not consistent with any prediction of the Lick/SDSS library even if  large systematic errors in their 
effective temperature estimates are assumed. In conclusion, the inconsistency of synthetic and observational indices for cool dwarfs  must be ascribed to 
inadequacies or incorrect assumptions in the models and/or synthetic spectra used to compute the Lick/SDSS library for $T_{\rm eff} < 5250$\,K.
In order to improve the theoretical predictions for these temperatures, we are now computing new atmosphere models and new synthetic spectra
to derive more reliable estimates of synthetic Lick/SDSS indices. The main differences between the new library and the published one 
will be:

\begin{enumerate}
 \item the use of new molecular opacities (in particular we now use a new release  of the H$_2$O line list made by Kurucz
from Partridge \& Schwenke\footnote{http://kurucz.harvard.edu/molecules/h2o/h2ofastfix.readme});
\item the computation of new atmosphere models with ATLAS12 code \cite{KU05} which allow us to adopt a microturbulence velocity of 1\,km\,s$^{-1}$ which is more 
appropriate for dwarfs than the 2\,km\,s$^{-1}$ of the Opacity Distribution Functions used in \cite{FR10};
\item  the use of a new updated version of the SPECTRUM code (i. e. v2.76f\footnote{http://www.appstate.edu/~grayro/spectrum/spectrum.html})  by \cite{GR94} to compute
synthetic spectra;
\item the use of a new revised line list containing TiO data from \cite{PL98}.
This line list (based on cool5.iso.lst, kindly provided to us by R. O. Gray 2010, private communication) 
will be characterized by the use of  empirical  log\,{\it gf} values for the strongest lines which we are deriving  
from the comparison between solar synthetic and observed high resolution spectra. The line list will include also improvements
from SpectroWeb \citep{LO08}. 
 \end{enumerate}

Preliminary results are shown in Fig.\,\ref{newteff} for CN$_2$, Mgb, and NaD: the new synthetic index values are in much better agreement
with the observational ones than the original  Lick/SDSS library predictions. These results show  that significant improvements can be achieved 
with the above described updated tools and suggest that the already very good reliability of the Lick/SDSS library for $T_{\rm eff} > 5250$\,K 
could be extended also for lower temperatures in the forthcoming new version of the library (\citealt{FR14}).

\begin{figure*}
\includegraphics[]{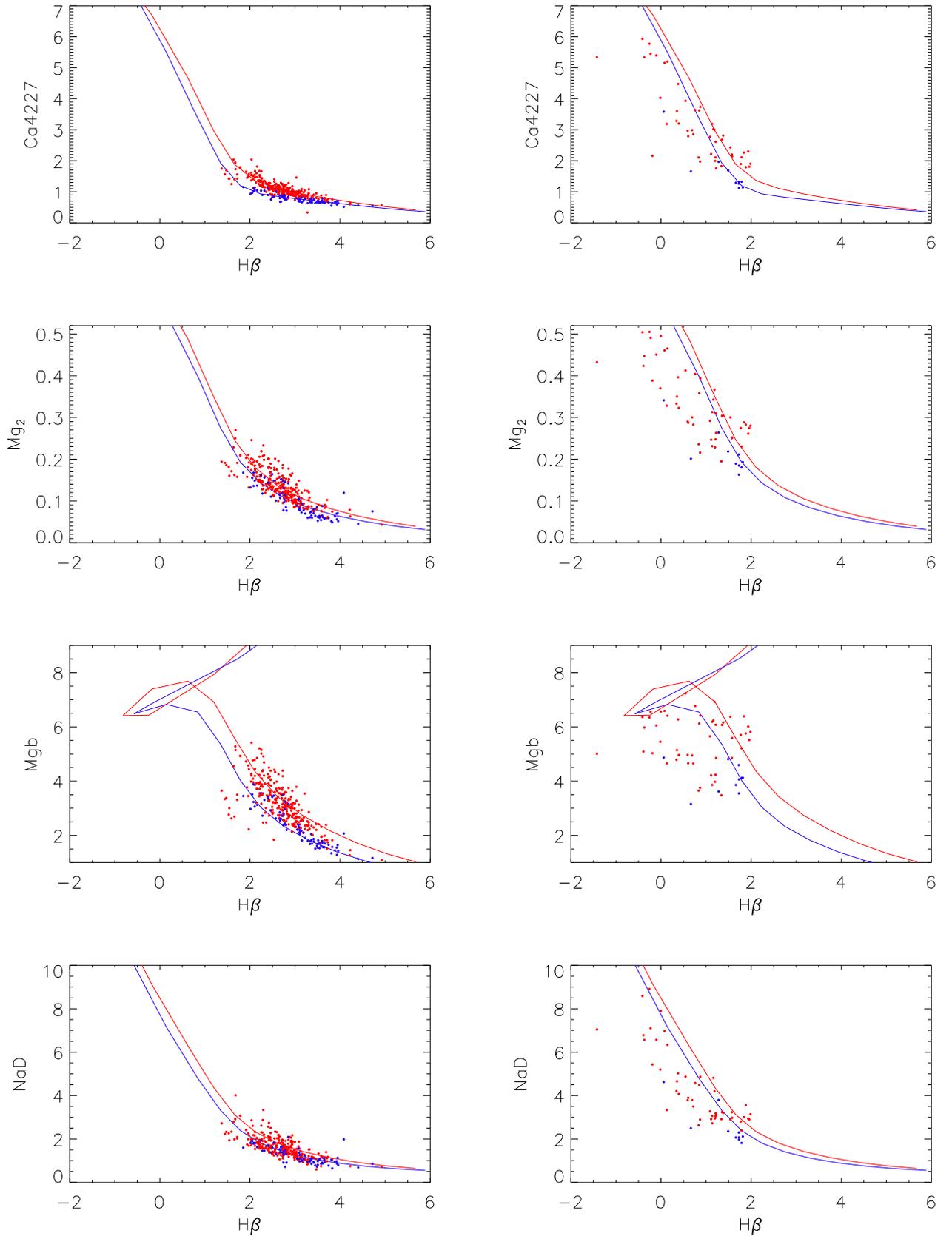}
\caption{Index--index diagrams: FEROS--Lick/SDSS indices of {\it dwarfs} (red points) and  {\it subgiants} (blue points) with 
$T_{\rm eff}\le 5250$\,K (right panels) or  $T_{\rm eff} > 5250$\,K (left panels)  are superimposed onto the predictions of the 
Lick/SDSS library at [Fe/H]=+0.0, [$\alpha$/Fe]=+0.0, and log\,$g$=4.0 (blue lines) or  log\,$g$=4.5 (red lines).}
\label{ind}
\end{figure*}

\clearpage
\begin{figure}
\includegraphics[width=80mm]{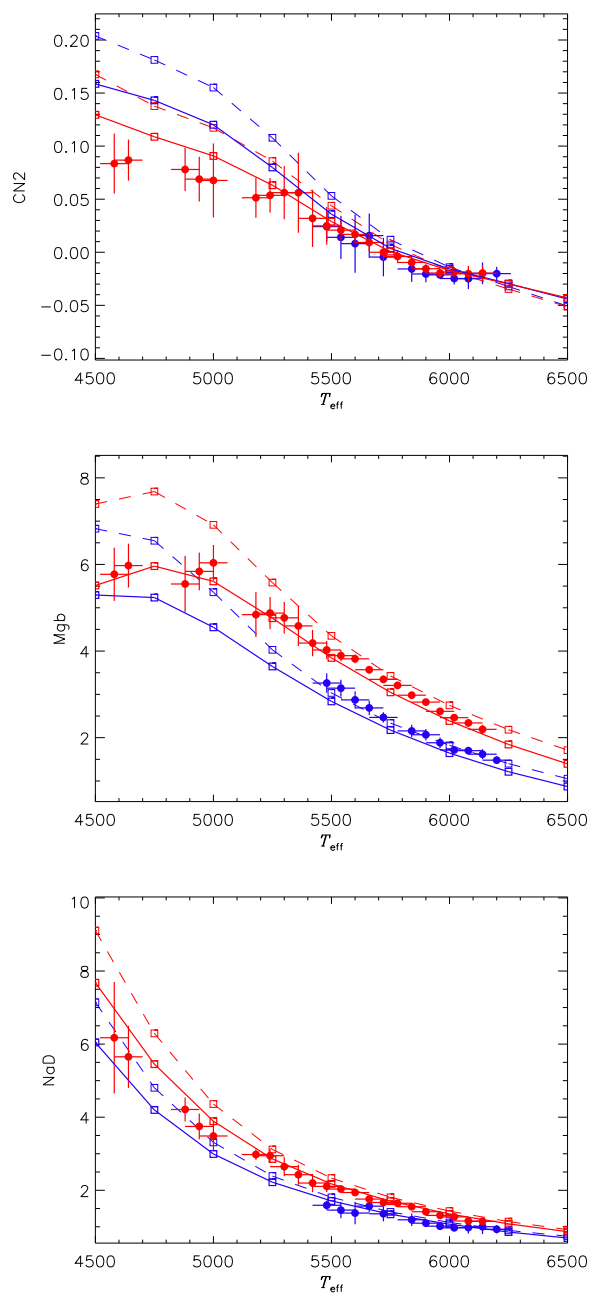}
\caption{Preliminary results of computation of new synthetic Lick/SDSS index values at [Fe/H]=+0.0, [$\alpha$/Fe]=+0.0, 
and log\,$g$=4.0 (blue solid line) and  log\,$g$=4.5 (red solid line). The comparison with
FEROS--Lick/SDSS observational  indices for {\it dwarfs} (red points) and  {\it subgiants} (blue points) is illustrated: 
observational data
are re--binned as in Fig.\ref{teff} and  predictions of the original Lick/SDSS library (dashed lines) are also shown as  reference. }
\label{newteff}
\end{figure}

\section*{Acknowledgments}
Based on data obtained from the ESO Science Archive Facility under request numbers 88412 and 92380. This work received partial financial support
from the Mexican CONACyT via grant SEP-2009-134985  and from PRIN MIUR 2010--2011 project ``The Chemical and dynamical Evolution of the Milky Way
and Local Group Galaxies'', prot. 2010LY5N2T.

\bsp

\label{lastpage}

\end{document}